# The Momentum Representation of the Hydrogen Atom in Paraboloidal Coordinates

John R. Lombardi
Department of Chemistry
The City College of New York
New York, N.Y. 10031

We examine the procedure to construct the variables of use for the momentum representation in quantum mechanics. The momentum variables must be chosen properly conjugate to the corresponding position space variables, such that valid uncertainty relationships are maintained. We then apply such considerations to the hydrogen atom to obtain the momentum space wave functions corresponding to the position space functions in paraboloidal coordinates. The advantages and disadvantages of employing the momentum representation are explored.

**The Momentum Representation**

In quantum mechanics it is well-known that the state functions, which are eigenfunctions of the Hamiltonian operator, may be regarded solely as functions of the position of the particles $q$. The position variable has a corresponding variable in momentum $p$, to which it is conjugate. Alternatively, the state functions can be represented as functions solely of the momentum. Both representations are equivalent, and the expectation values of quantum mechanical observables should be the same, independent of the representation chosen. The selection of a particular representation is entirely a matter of convenience, and as a result it is usually chosen to be the position space representation. This is because, at least for bound-state problems, it is easier to express the potential energy (usually a function of position variables) in the position representation. By contrast, in the momentum representation the position variables are represented by differential or integral operators. Consequently, little work has been carried out for bound-state problems in momentum space. This is unfortunate, since in many ways the momentum representation may be thought of as the other half of quantum mechanics. This alone, makes it worthwhile at least to examine and investigate the solutions for important problems in momentum space. More importantly, there are sometimes distinct practical advantages to momentum space functions. As an example, we show here that a correct interpretation of the momentum operator indicates that it must be regarded as a complex variable[1]. Since many solutions to the integral eigenvalue equations obtained for atomic problems result in functions which have poles along the imaginary axis, powerful analytic tools such as Cauchy's theorem may be invoked in order to facilitate integration over these functions. A rather large stumbling block in the calculation of accurate wave functions for polyatomic molecules is the extensive number of integrals which must be calculated, often numerically. Use of momentum space basis sets may alleviate this problem.

The momentum representation in quantum mechanics may be defined as the space of momentum coordinates chosen *properly conjugate* to each of the corresponding position space coordinates. Most quantum mechanical calculations begin with the establishment of an appropriate classical Hamiltonian written in terms of the position $q$ and momentum $p$ of the particles involved. In Cartesian space the variables $p, q$ range from $[-\infty,\infty]$. For a single particle in one dimension it is usually written as:

$$H(p,q) = p^2/2\mu + V(q) \qquad (1)$$

Where $\mu$ is the mass of the particle and $V(q)$ is the potential energy function. In order to convert this classical Hamiltonian function to a quantum mechanical operator, we express the momentum as an operator (using units in which $\hbar =1$) as

$$p \rightarrow -id/dq \qquad (2)$$

and we seek solutions of the resulting differential eigenvalue equation:

$$\mathbf{H}\psi(q) = E\psi(q) \qquad (3)$$

$E$ represents the eigenvalues, while $\psi(q)$ are the state eigenfunctions, expressed as functions of the position of the particle. It is easy to show that a commutation expression between the conjugate operators can be written:

$$[p,q] = i \qquad (4)$$

State functions expressed as functions of position $q$, are usually referred to as the position representation. An equally valid description of the state of the system is through functions of the momentum $\varphi(p)$, which are related to the position functions by way of a unitary transform[2] $S(p,q)$ such that:

$$\phi(p) = (2\pi)^{-1} \int_{-\infty}^{\infty} S(p,q)\psi(q)dq \qquad (5)$$

To determine the correct equation leading to $\varphi(p)$, we replace the functions of $q$ in (1) with operators:

$$q \to id/dp, \qquad q^{-1} \to -i\int dp \equiv \mathbf{I} \qquad (6)$$

Note expression given for $q^{-1}$ is shorthand; the path of integration for the expression for $q^{-1}$ is along the real axis from $-\infty$ to $p$. The complete expression for $\mathbf{I} f(p)$ is therefore $-i \int_{-\infty}^{p} f(p')dp'$. It should also be pointed out that Hylleras[3] discusses the meaning of the operator form of the inverse position.

Depending on the form of $V(q)$ we obtain either differential or integral equations, which must be solved for $\varphi(p)$. Functions in this form are usually referred to as the momentum representation. As noted previously, most quantum mechanical calculations are carried out in the position representation, due to the fact that it is much easier to determine the potential energy $V$ in terms of position $q$. However, for completeness it is worthwhile to find the correct expressions for state functions in terms of $p$ as well. In fact, in some cases, this leads us to possible simplifying advantages as we will see below.

It is also worthwhile to determine the correct transform between position representation functions and those in the momentum representation. It is often stated in textbooks that this transform is the Fourier transform.[4] In this case we can write simply:

$$S(p,q) = e^{iqp} \qquad (7)$$

However, as was first pointed out by Podolsky[5], this is correct only for Cartesian coordinates. For curvilinear coordinates, we must choose a transform which maintains the proper conjugate relation between momentum space and position space variables. DeWitt[6] obtained a more general form, which holds for all transformations between a *n*-dimensional Cartesian space ($x_1, x_2, ...x_n$) and an arbitrary coordinate system ($q_1, q_2...q_n$) such that the volume element $dV = dx_1 dx_2...dx_n$ is transformed into $dV = g dq_1 dq_2...dq_n$. The function $g$ is the Jacobian of transformation.[7] DeWitt then showed that the Hermetian form of the momentum space variables is:

$$p_k \to -i\left(\frac{\partial}{\partial q_k} + \frac{1}{2g}\frac{\partial g}{\partial q_k}\right) \qquad (8)$$

Under these circumstances, the most general transform between position space and momentum space is:

$$\phi(p_1...p_n) = (2\pi)^{-1}\int_{-\infty}^{\infty} S(p_1...p_n, q_1...q_n)\psi(q_1...q_k) g\, dq_1...dq_n \qquad (10)$$

$$S(p_1...p_n, q_1...q_n) = \frac{1}{\sqrt{g}} e^{i\vec{q}\cdot\vec{p}} \qquad (11)$$

Early work was carried out on the hydrogen atom by Podolsky and Pauling[8] in which the momentum distribution was obtained as the square of the momentum eigenfunction. This was obtained by a direct Fourier transform of the position space functions in terms of the variables *(P, Θ, Φ)*, where *P* is the total momentum, and *Θ, Φ* are the polar angle of *P* measured from the position space axes. In contrast, Fock[9] used an integral equation approach with the same variables to obtain identical results. These variables were not chosen to be conjugate with any position space variables. In fact the angular parts are defined in terms of position space, and in this regard the eigenfunctions presented cannot be considered to be in true momentum representation. No operator expression for the momentum variables is given, nor are there any commutation relations. This is in apparent contradiction to the earlier work by Podolsky[5] in which the correct curvilinear momentum expressions were derived.

In previous work[1] we have obtained the momentum representation functions for the hydrogen atom using spherical polar coordinates. For the readers convenience, a summary of this derivation is presented in the appendix of this work [note this also includes a corrected normalization constant]. This has proven valuable for various extensions of momentum representation to the He atom[10,11], relativistic Dirac equation[12], as well as the hydrogen molecule ion[13].

In this work, we explore the extension of these ideas to the momentum representation of the hydrogen atom in paraboloidal coordinates. A valuable description of this topic in position space is given by Bethe and Salpeter[2]. As they point out, this coordinate system is useful in considering various important perturbation problems, such as the Stark effect, in which an external field is applied to the atom, the photo-electric effect, the Compton effect, as well as for electron collisions. We should add that use of these coordinates may also be of value in considering molecular problems where there are other centers of positive charge, such as diatomic molecules.

**Hydrogen Atom in Paraboloidal Coordinates**

Paraboloidal coordinates are especially useful for perturbation calculations in which a particular direction in space is preferred by some external force. We use the position space formulation of Bethe and Salpeter[2] (p. 27) in which we transform from Cartesian ($x,y,z$) to paraboloidal ($u,v,\varphi$) coordinates using:

$$x = \sqrt{uv}\, \cos\varphi$$

$$y = \sqrt{uv}\, \sin\varphi$$

$$z = 1/2(u - v)$$

$$r = 1/2(u + v) \quad (12)$$

Letting: $Z = Z_1 + Z_2$, after separation of variables in position space we obtain the three equations:

$$\frac{1}{u}\frac{\partial}{\partial u}u\frac{\partial U}{\partial u} + \left[\frac{1}{2}E + \frac{Z_1 e^2}{u} - \frac{m^2}{4u^2}\right]U(u) = 0$$

$$\frac{1}{v}\frac{\partial}{\partial v}v\frac{\partial V}{\partial v} + \left[\frac{1}{2}E + \frac{Z_2 e^2}{v} - \frac{m^2}{4v^2}\right]V(v) = 0$$

$$\frac{\partial^2 \Phi}{\partial \phi^2} = -m^2 \Phi \qquad (13)$$

In this coordinate system, the Jacobean of transformation is $g(u,v) = \frac{1}{4}(u+v)^2$ which is not separable into a product of one-dimensional terms. Consequently, we must start with the separated position space equations and transform them to momentum space individually. Following Dirac[14] (section 138), in order to preserve a proper conjugate relation between position and momentum operators, we require the commutators: $[u,p_u] = i$ and $[v,p_v] = i$. Then, in momentum space the relevant operators become:

$$p_u \rightarrow -i\left(\frac{d}{du} + \frac{1}{2u}\right)$$

$$p_v \rightarrow -i\left(\frac{d}{dv} + \frac{1}{2v}\right)$$

$$p_\varphi \rightarrow -i\frac{d}{d\varphi} \qquad (14)$$

$$\frac{1}{u} \rightarrow -i\int dp_u$$

$$\frac{1}{v} \rightarrow -i\int dp_v$$

The momentum equations are then

$$\left[p_u^2 - \mu E + \frac{m^2-1}{4}i^2 \int dp_u \int dp_u + 2\mu Z_1 e^2 i \int dp_u\right] U(p_u) = 0 \qquad (15)$$

$$\left[p_v^2 - \mu E + \frac{m^2-1}{4}i^2 \int dp_v \int dp_v + 2\mu Z_2 e^2 i \int dp_v\right] V(p_v) = 0$$

$$[p_\varphi^2 - m^2]\rho(p_\varphi) = 0$$

The solution to the third $p_\phi$ equation is simply: $\rho(p_\phi) = \delta(p_\phi \pm m)$. Here $m$ takes on integer values including zero. Note the other two equations are identical to each other and to the solutions in the appendix equation (A6) with the adjustments that $2\mu E$ has been replaced by $\mu E$, and $\ell(\ell+1)$ replaced by $(m^2-1)/4$ such that m = $2\ell +1$. Here $\ell$ (which takes on ½ integer values for $m$ even) does not have the same meaning as in spherical polar coordinates, but should be regarded solely as a dummy index, used in obtaining the correct form of α. We also introduce the quantum numbers $n_1$, and $n_2$ such that $n_i + \frac{1}{2}(m+1) = nZ_i/Z$. The two equations are identical to each other and are the same as the radial momentum expression in spherical polar coordinates (See appendix eqn A6 or ref 1), so we may solve them simultaneously, taking: $p = p_u = p_v = p_r$ and letting either $U$ or $V$ be replaced by $a_{n,n_1\ell}$ or $a_{n,n_2\ell}$ respectively.

$$\left[p^2 - \mu E + \frac{m^2-1}{4}i^2 \int dp \int dp + 2\mu Z_i e^2 i \int dp\right] \alpha(p) = 0 \qquad (16)$$

After integrating and simplifying we obtain[2] the normalized expression for α(p):

$$a_{n,n_i\ell} = N_{n,n_i,\ell} \sum_{k=0}^{E(n_i-\ell-1)} \left(\frac{2^k(n_i-\ell-k)_k}{k!(\ell+k+2)_\ell}\right) \left(\frac{\frac{ip_0}{n}}{p-\frac{ip_0}{n}}\right)^{\ell+k+2} \qquad (17)$$

Where:

$$N_{n,n_i\,\ell} = \frac{2^{\ell+1}}{\Gamma(n_i-\ell)\sqrt{\pi s_{n,i\ell}}} \left(\frac{n}{p_0}\right)^{1/2}$$

$$S_{n,n_i,\ell} = \sum_{k=0}^{E(n_i-\ell-1)} \sum_{j=0}^{E(n_i-\ell-1)} \frac{(-1)^{k+j}(2\ell+k+j+2)!}{k!j!\Gamma(ni-\ell-k)\Gamma(n_i-\ell-j)(2\ell+k+1)!(2\ell+j+1)!} \quad (18)$$

These remarkable results illustrate several interesting aspects to the momentum representation. First, note that either momentum variable $p$ ($p_u$ or $p_v$) must be considered as a complex variable, despite the fact that all integrals of quantum mechanical interest are along the real axis, in the domain $[-\infty,\infty]$. Note $p_0 = \mu Z e^2 = 1/a_0$, and thus may be termed the "Bohr momentum" by analogy with the Bohr radius ($a_0$).

It may be observed that although this expression appears somewhat complicated, it involves a simple sum over poles (for integer $\ell$) or branch line (for ½ integer $\ell$)[15] of various orders located in the upper half plane along the imaginary axis at $ip_0/n$. The functions $\alpha_{n,ni,\ell}(p)$ themselves are complex as well. We can now suggest a very simple way to determine integrals, especially those involving expectation values of various operators. We invoke Cauchy's Theorem:[15,16]

$$\oint \frac{f(z)dz}{(z-a)^{n+1}} = \frac{2\pi i}{n!} f^n(a)$$

where the integral is over a closed path enclosing the pole (n = integer) or branch line (for n = ½ integer), and $f^n(a)$ is the $n^{th}$ derivative of $f(z)$ evaluated at the pole $z = a$. Many integrals which are difficult to obtain in position space (such as those for high values of $n$) can be obtained by inspection using this expression.

The complete momentum eigenfunction for the hydrogen atom in paraboloidal coordinates is:

$$\emptyset_{n_1 n_2 m}(p_u, p_v, p_\varphi) = \alpha_{n,n_1,m}(p_u)\alpha_{n,n_2,m}(p_v)\delta(p_\varphi \pm m) \quad (19)$$

In this expression the functions α are those given in equation (17). This expression emphasizes the symmetry between $p_u$ and $p_v$. Letting $n = n_1 + n_2 + m + 1$, the energy may be expressed by:

$$E = -\frac{Z^2 \mu e^4}{2n^2} = -\frac{Z^2 \mu e^4}{2(n_1+n_2+m+1)^2} \tag{20}$$

In this expression, $n_1$ and $n_2$ are non-negative integers such that $n_{1,2} = 0,1,2...n-m-1$. Since for m>0 there are two possible values (±m), the correct degeneracy is obtained.

As examples of the simplicity of these functions we present the normalized momentum eigenfunction for the $n = 1, 2$ states of the hydrogen atom:

$$\emptyset_{0,0,0}(p_u, p_v, p_\varphi) = \frac{-ip_0^2}{2(p_u-ip_0)^{3/2}(p_v-ip_0)^{3/2}} \delta(p_\varphi) \tag{21}$$

$$\emptyset_{1,0,0}(p_u, p_v, p_\varphi) = \frac{ip_0^2}{4\sqrt{2}} \frac{(p_u+ip_0/4)}{(p_u - ip_0/2)^{5/2}(p_v - ip_0/2)^{3/2}} \delta(p_\varphi)$$

$$\emptyset_{0,1,0}(p_u, p_v, p_\varphi) = \frac{ip_0^2}{4\sqrt{2}} \frac{(p_v + ip_0/4)}{(p_u - ip_0/2)^{3/2}(p_v - ip_0/2)^{5/2}} \delta(p_\varphi)$$

$$\emptyset_{0,0,1}(p_u, p_v, p_\varphi) = \frac{p_0^3}{4\pi} \frac{1}{(p_u - ip_0/2)^2(p_v - ip_0/2)^2} \delta(p_\varphi \pm 1)$$

Three dimensional plots of the real parts of each of these functions are shown in figures (1)-(4). (The functions $\delta(p_\varphi \pm m)$ are not included.)

$$\emptyset_{0,0,0}(p_u, p_v, p_\varphi) = \frac{-ip_0^2}{2(p_u - ip_0)^{3/2}(p_v - ip_0)^{3/2}} \delta(p_\varphi)$$

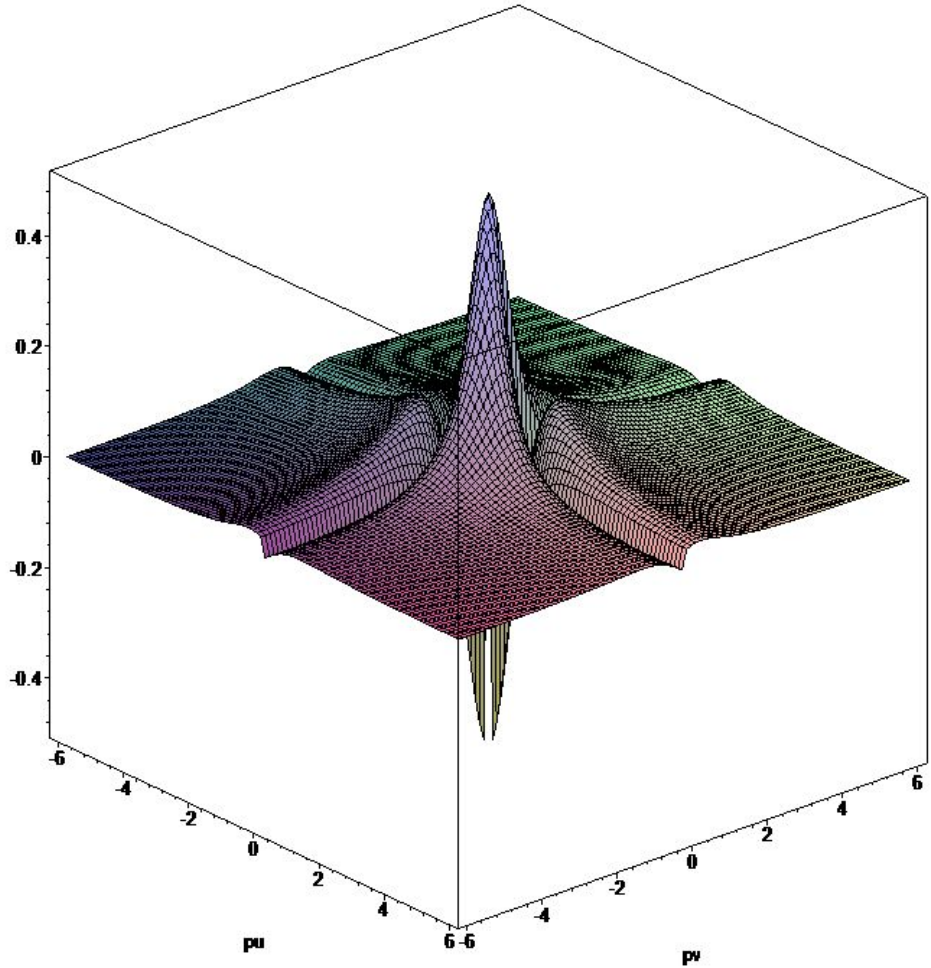

Figure (1). The real part of the function $\emptyset_{0,0,0}$ along the $p_u$ and $p_v$ axes.

$$\emptyset_{1,0,0}(p_u, p_v, p_\varphi) = \frac{ip_0^2}{4\sqrt{2}} \frac{(p_u + ip_0/4)}{(p_u - ip_0/2)^{5/2}(p_v - ip_0/2)^{3/2}} \delta(p_\varphi)$$

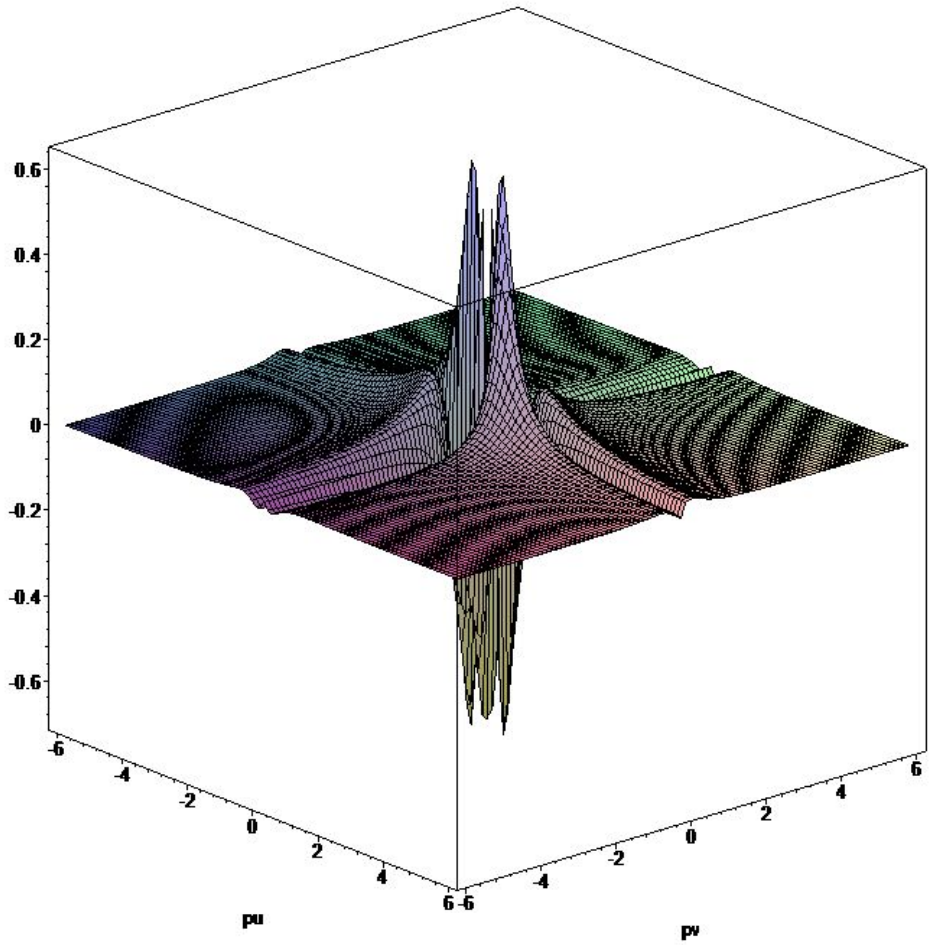

Figure (2). The real part of the function $\emptyset_{1,0,0}$ along the $p_u$ and $p_v$ axes.

$$\emptyset_{0,1,0}(p_u, p_v, p_\varphi) = \frac{ip_0^2}{4\sqrt{2}} \frac{(p_v + ip_0/4)}{(p_u - ip_0/2)^{3/2}(p_v - ip_0/2)^{5/2}} \delta(p_\varphi)$$

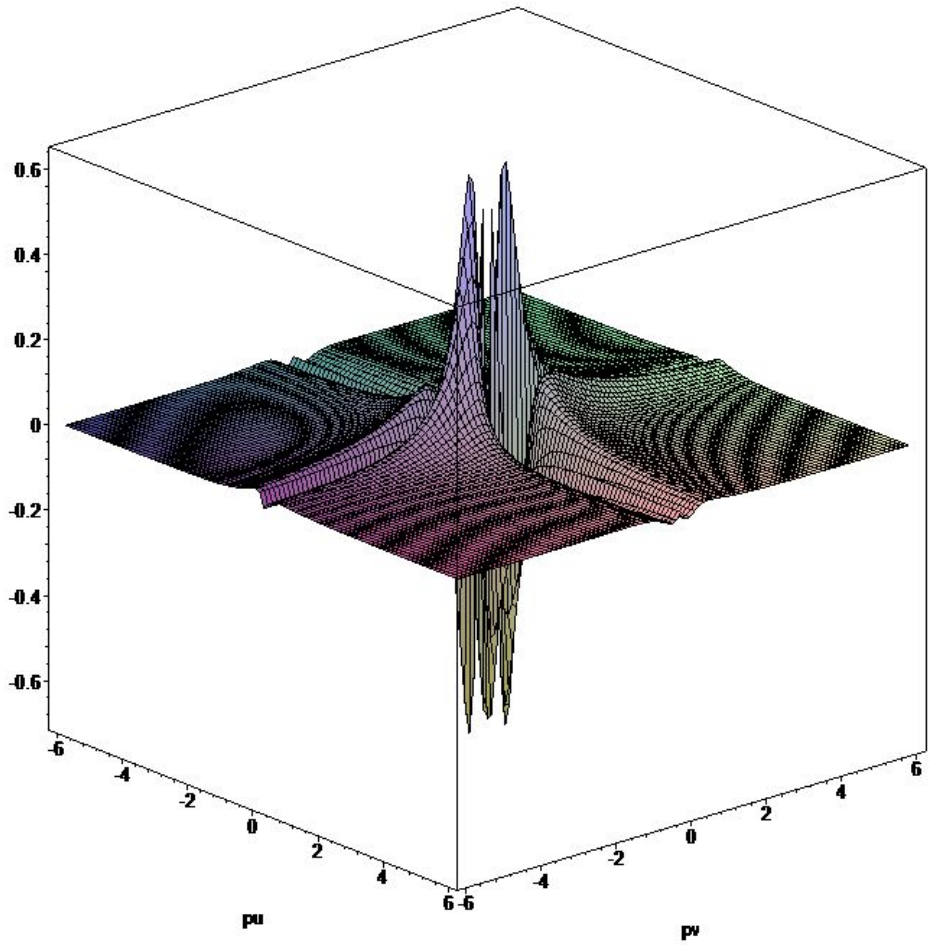

Figure (3). The real part of the function $\emptyset_{0,1,0}$ along the $p_u$ and $p_v$ axes.

$$\emptyset_{0,0,1}(p_u, p_v, p_\varphi) = \frac{p_0^3}{4\pi} \frac{1}{(p_u - ip_0/2)^2 (p_v - ip_0/2)^2} \delta(p_\varphi \pm 1)$$

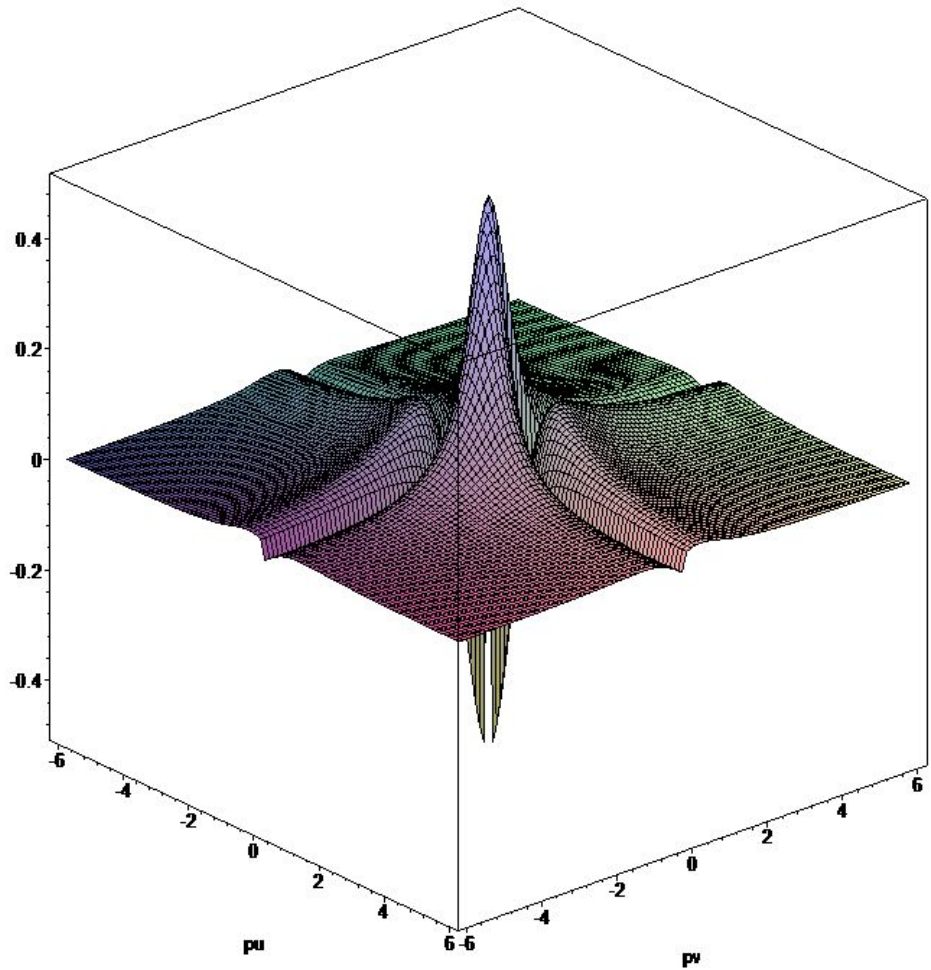

Figure (4). The real part of the function $\emptyset_{0,0,1}$ along the $p_u$ and $p_v$ axes. Note this is identical to $\emptyset_{0,0,-1}$

It is also of interest to compare these results to those of the appendix, namely equation (A28) for the functions in momentum space for spherical polar coordinates. Note that the present functions are polar in both $p_u$ and $p_v$ directions, and do not involve Bessel functions. In the light of Cauchy's theorem, this represents considerable computational advantages over the spherical polar results.

**Conclusions**

We have examined the correct expression for momentum variables which may be chosen properly conjugate to corresponding position variables. These must be chosen so that proper uncertainty relations are also obtained. The resulting expressions allow the construction of eigenfunction equations in momentum space, which may involve either differential or integral equations depending on the exact form of the potential energy. Solution of these equations provides momentum representation eigenfunctions, which may be used to obtain expectation values of all the operators of interest.

We illustrate the advantages of this procedure by application to the hydrogen atom paraboloidal coordinates. The resulting integral equations may readily be solved, resulting in eigenfunctions which involve extension of the momentum variables to complex space. These functions can then be represented by simple polar functions with singularities along the imaginary *p ($p_u$ or $p_v$)* axis. These poles enable application of Cauchy's theorem for determination of various integrals, leading to facile calculations of integral which are of considerably more difficulty in position space.

**Acknowledgements**

The author would like to thank Dr. J. F. Ogilvie for encouragement and generous help in checking the calculations using Maple software, not to mention cogent advice as well. He also provided the three-dimensional plots of the momentum functions shown here. Important assistance in generalizing the β functions (equations A24 and A25) was provided by Dr. Brian Thomas. Further help was provided by Gabriella Clemente and Catherine Abrams. Financial support was provided by the City University of New York, and the National Science Foundation (CHE-1402750).

**Appendix: The Hydrogen atom in momentum space; Spherical Polar coordinates**

**Note:** This appendix is presented for the convenience of the reader. For more detailed derivation, see ref 1. Note, however, that ref. 1 is slightly updated in equations (A16) and (A17), as well as (A24) and (A25).

**The Hydrogen Atom-—Radial Momentum Functions**

We consider three momentum variables ($p_r$, $p_\theta$, $p_\varphi$), which must be chosen as conjugates to proper position spherical polar variables ($r$, $\cos\theta$, $e^{i\varphi}$). Since the hydrogen atom is separable in the position representation, we may transform the radial equation and the angular equations separately. The radial eigenfunctions in momentum representation will be called $\alpha(p_r)$. Remembering that the angular eigenfunctions of the angular momentum operators ($L^2$ and $L_z$) in position space are the spherical harmonics $Y_{l,m}(\theta,\varphi)$, with eigenvalues $\ell(\ell+1)$ and $m$, we shall call the corresponding angular eigenfunctions in the momentum representation $\beta(p_\theta)\rho(p_\varphi)$. They satisfy the expression $L^2\beta(p_\theta)\rho(p_\varphi) = \ell(\ell+1)\beta(p_\theta)\rho(p_\varphi)$ and $L_z\rho(p_\varphi) = m\rho(p_\varphi)$. The total momentum wave functions for the hydrogen atom in momentum space will then be written:

$$\phi(p_r, p_\theta, p_\phi) = \alpha(p_r)\beta(p_\theta)\rho(p_\phi) \tag{A1}$$

We first turn to the examination of momentum space variable $p_r$, which is conjugate to the position space radial variable $r$. As suggested above, we obtain the conjugate momentum variable to be expressed by: [6]

$$p_r \rightarrow -i\left(\frac{\partial}{\partial r} + \frac{1}{r}\right) \tag{A2}$$

We choose expressions for the position space variables to be

$$r \to i\frac{\partial}{\partial p_r} \qquad\qquad r^{-1} \to -i\int dp_r \equiv I \qquad\qquad (A3)$$

We then obtain the following momentum representation equation:

$$\left[\left(p_r^2 - 2\mu E\right) + \ell(\ell+1)I^2 + 2\mu Z e^2 I\right]\alpha(p_r) = 0 \qquad\qquad (A4)$$

This is a second order integral equation in $p_r$ which must be solved for $\alpha(p_r)$.

**s-states ( $\ell = 0$)**

We first solve this equation for the important case for which $\ell = 0$. Then we have simply:

$$\left[\left(p_r^2 - 2\mu E\right) + 2\mu Z e^2 i \int dp_r\right]\alpha(p_r) = 0 \qquad\qquad (A5)$$

It is convenient to make the substitution $p_0 = \mu Z e^2 = 1/a_0$ (where $a_0$ is the Bohr Radius). Taking the first derivative of this equation and rearranging we obtain:

$$\frac{d\alpha(p_r)}{dp_r} = -\frac{2p_r + 2p_0 i}{p_r^2 - 2\mu E}\alpha(p_r) \qquad\qquad (A6)$$

This may be integrated easily to obtain:

$$\alpha(p_r) = \frac{N}{p_r^2 - 2\mu E}\left(\frac{p_r - \sqrt{2\mu E}}{p_r + \sqrt{2\mu E}}\right)^{\frac{ip_0}{\sqrt{2\mu E}}} \qquad\qquad (A7)$$

Since $\alpha(p_r)$ must be single-valued we require:

$$\frac{ip_0}{\sqrt{2\mu E}} = n \qquad\qquad (A8)$$

where $n$ is a positive integer. Solving for $E$ we obtain the correct energy for the hydrogen atom:

$$E = -Z^2\mu e^4/2n^2 \qquad (A9)$$

Normalizing, we obtain the simple result:

$$\alpha_{n,0}(p_r) = \sqrt{\frac{2p_0^3}{\pi n^3}} \frac{(p_r + ip_0/n)^{n-1}}{(p_r - ip_0/n)^{n+1}} \qquad (A10)$$

Note that the ground state of the hydrogen atom *(n=1)* is still simpler in form:

$$\alpha_{1,0}(p_r) = \sqrt{\frac{2p_0^3}{\pi}} \frac{1}{(p_r - ip_0)^2} \qquad (A11)$$

**The general solution for any $\ell$.**

We now examine the solutions of equation (A4) for any value of $\ell$. Note this involves a second order integral ($I^2$) equation. This may be carried out most easily by taking the ($\ell+1$)$^{th}$ derivative of the equation. Then we obtain:

$$\frac{d\chi(p_r)}{dp_r} = -\frac{2(\ell+1)p_r + 2ip_0}{p_r^2 - 2\mu E}\chi(p_r) \qquad (A12)$$

Where we have defined:

$$\chi(p_r) = i^\ell \frac{d^\ell}{dp_r^\ell}\alpha(p_r) \qquad (A13)$$

After integrating and simplifying we obtain[2] the normalized expression for $\alpha(p_r)$:

$$a_{n,\ell} = N_{n,\ell} \sum_{k=0}^{n-\ell-1} \binom{n-\ell-1}{k} \left(\frac{2^k(\ell+k+1)!}{(2\ell+k+1)!}\right) \left(\frac{\frac{ip_0}{n}}{p - \frac{ip_0}{n}}\right)^{\ell+k+2} \qquad (A14)$$

$$N_{n,\ell} = \frac{2^{\ell+1}}{(n-\ell-1)!\sqrt{\pi s_{n,\ell}}} \left(\frac{n}{p_0}\right)^{1/2}$$

$$s_{n,\ell} = \sum_{k=0}^{n-\ell-1}\sum_{j=0}^{n-\ell-1} \frac{(-1)^{k+j}(2\ell+k+j+2)!}{k!\,j!\,(n-\ell-k-1)!(n-\ell-j-1)!(2\ell+k+1)!(2\ell+j+1)!} \qquad (A15)$$

It may be observed that although this expression appears somewhat complicated, it involves a simple sum over poles of various orders ($\ell+2$ to $n$) located in the upper half plane along the imaginary axis at $ip_0/n$. Thus, once again, Cauchy's theorem can be invoked to obtain integrals over this function in a very simple manner. (Note this represents a correction to the normalization constant given in ref. 1)

### Angular Functions in Momentum Space

We now turn our attention to the momentum analogs of the position space angular functions, often called the spherical harmonics $Y_\ell^m(\theta,\phi)$. We may regard the variable $p_\theta$ to be conjugate to $u$ (= $\cos\theta$) and $p_\varphi$ conjugate to $v$ (= $e^{i\varphi}$) instead of ($\theta, \varphi$). Thus, we consider the variables ($p_r, p_\theta, p_\varphi$) to be chosen conjugate to the set of variables ($r, u, v$). This, of course, makes no difference to the well-known position space results, but ensures that our momentum space variables have a meaningful uncertainty relationship with the proper position space variables. We now can write:

$$p_\theta = -i\frac{\partial}{\partial u} \quad \text{and} \quad p_\phi = -iv\frac{\partial}{\partial v} \qquad (A16)$$

$$u = i\frac{\partial}{\partial p_\theta} \quad \text{and} \quad v = i\frac{\partial}{\partial p_\phi} \qquad (A17)$$

Remembering that the angular eigenfunctions in the momentum representation are $\beta(p_\theta)\rho(p_\varphi)$, we have the transformed momentum space equations:

$$L^2\beta(p_\theta)\rho(p_\phi) = \ell(\ell+1)\beta(p_\theta)\rho(p_\phi) \tag{A18}$$

Separating variables, we obtain an equation for ρ:

$$\left(v^2 \frac{\partial^2}{\partial v^2} + v\frac{\partial}{\partial v}\right)\rho(p_\phi) \rightarrow p_\phi^2 \rho(p_\phi) = m^2 \rho(p_\phi)$$
$$\left(p_\phi^2 - m^2\right)\rho(p_\phi) = 0 \tag{A19}$$

We have chosen the constant of separation to be $m^2$. This equation has the simple solution:

$$\rho_m(p_\phi) = \delta(p_\phi \pm m) \tag{A20}$$

Here $\delta$ is the Dirac delta function, and $m$ must be an integer (positive, negative or zero).

Utilizing the well-known properties of Bessel Functions[17] we can show that solutions to the equation for $\beta(p_\theta)$:

For $\ell - |m| = 2n$ \hfill (A21)

$$\beta_l^m(p_\theta) = \frac{N_\beta 2^{|m|} \Gamma\left(\frac{1}{2}\right)\Gamma\left(\frac{|m|}{2}+1\right)\Gamma\left(\frac{|m|+1}{2}\right)}{\Gamma\left(\frac{1}{2}-\frac{l+|m|}{2}\right)\Gamma\left(1+\frac{l-|m|}{2}\right)\Gamma\left(\frac{|m|}{2}+\frac{3}{2}+n\right)}\left(\frac{2}{p_\theta}\right)^{\frac{|m|}{2}+\frac{1}{2}}$$

$$X \sum_{k=0}^{n} (2k + \frac{|m|}{2} + \frac{1}{2})(-i)^{2k} J_{\frac{|m|}{2}+2k+\frac{1}{2}}(p_\theta) \frac{(-n)_k (n + |m| + \frac{1}{2})_k}{k!\left(\frac{1}{2}\right)_k}$$

$$X \frac{(-1)^{n-k}(\frac{|m|}{2})_{n-k}(\frac{|m|}{2}+\frac{1}{2})_k(\frac{|m|}{2}+1)_k}{(\frac{|m|}{2}+\frac{3}{2}+n)_k}$$

For $\ell - |m| = 2n + 1$ \hfill (A22)

$$\beta_l^m(p_\theta) = \frac{-N_\beta 2^{|m|+1}\,\Gamma\!\left(\frac{1}{2}\right)\Gamma\!\left(\frac{|m|}{2}+1\right)\Gamma\!\left(\frac{|m|+3}{2}\right)}{\Gamma\!\left(\frac{1}{2}+\frac{l-|m|}{2}\right)\Gamma\!\left(-\frac{l+|m|}{2}\right)\Gamma\!\left(\frac{|m|}{2}+\frac{5}{2}+n\right)}\left(\frac{2}{p_\theta}\right)^{\frac{|m|}{2}+\frac{1}{2}}$$

$$X\sum_{k=0}^{n}(2k+\frac{|m|}{2}+\frac{3}{2})(-i)^{2k+1}J_{\frac{|m|}{2}+2k+\frac{3}{2}}(p_\theta)\frac{(-n)_k(n+|m|+\frac{3}{2})_k}{k!\,(\frac{3}{2})_k}$$

$$X\frac{(-1)^{n-k}(\frac{|m|}{2})_{n-k}(\frac{|m|}{2}+\frac{3}{2})_k(\frac{|m|}{2}+1)_k}{(\frac{|m|}{2}+\frac{5}{2}+n)_k}$$

Where:

$$N_\beta = \sqrt{\frac{i(2l+1)(l-|m|)!}{2(l+|m|)!}} \tag{A23}$$

and we use the Pochhammer symbol: $(a)_n = \Gamma(a+n)/\Gamma(a)$.

Note these solutions involve corrections to the normalization as well as a more complete expression of the angular functions $\beta(p_\theta)$ than those given in reference 1.

**Complete Hydrogenic Functions using Spherical Polar Coordinates**

We now present a few of the lowest (un-normalized) hydrogen functions in the momentum representation. Remembering that the total functions are products:

$$\phi_{n,\ell,m}(p_r,p_\theta,p_\phi) = \alpha_{n,\ell}(p_r)\beta_\ell^m(p_\theta)\rho_m(p_\phi) \tag{A27}$$

We obtain for $n = 1,2$. For plots of these functions, see figures A1-A4.

$$\phi_{1,0,0}(p_r,p_\theta,p_\phi) = \frac{1}{(p_r-ip_0)^2}\frac{1}{p_\theta^{1/2}}J_{1/2}(p_\theta)\delta(p_\phi) \tag{A28}$$

$$\phi_{2,0,0}(p_r,p_\theta,p_\phi) = \frac{(p_r+ip_0/2)}{(p_r-ip_0/2)^2}\frac{1}{p_\theta^{1/2}}J_{1/2}(p_\theta)\delta(p_\phi)$$

$$\phi_{2,1,0}(p_r, p_\theta, p_\phi) = \frac{1}{(p_r - ip_0/2)^3} \frac{1}{p_\theta^{1/2}} J_{3/2}(p_\theta)\delta(p_\phi)$$

$$\phi_{2,1,\pm1}(p_r, p_\theta, p_\phi) = \frac{1}{(p_r - ip_0/2)^3} \frac{1}{p_\theta} J_1(p_\theta)\delta(p_\phi \pm 1)$$

As noted previously, among the advantages of the momentum representation is the relative ease in integrating over these functions. All expectation values involve such integrals. In addition to the above-mentioned ability to use Cauchy's theorem to integrate over the radial momentum variable, note also that integrating over $p_\varphi$ can usually be carried out by inspection. The only difficulties involve integration over $p_\theta$.

We now present several three-dimensional plots of these functions.

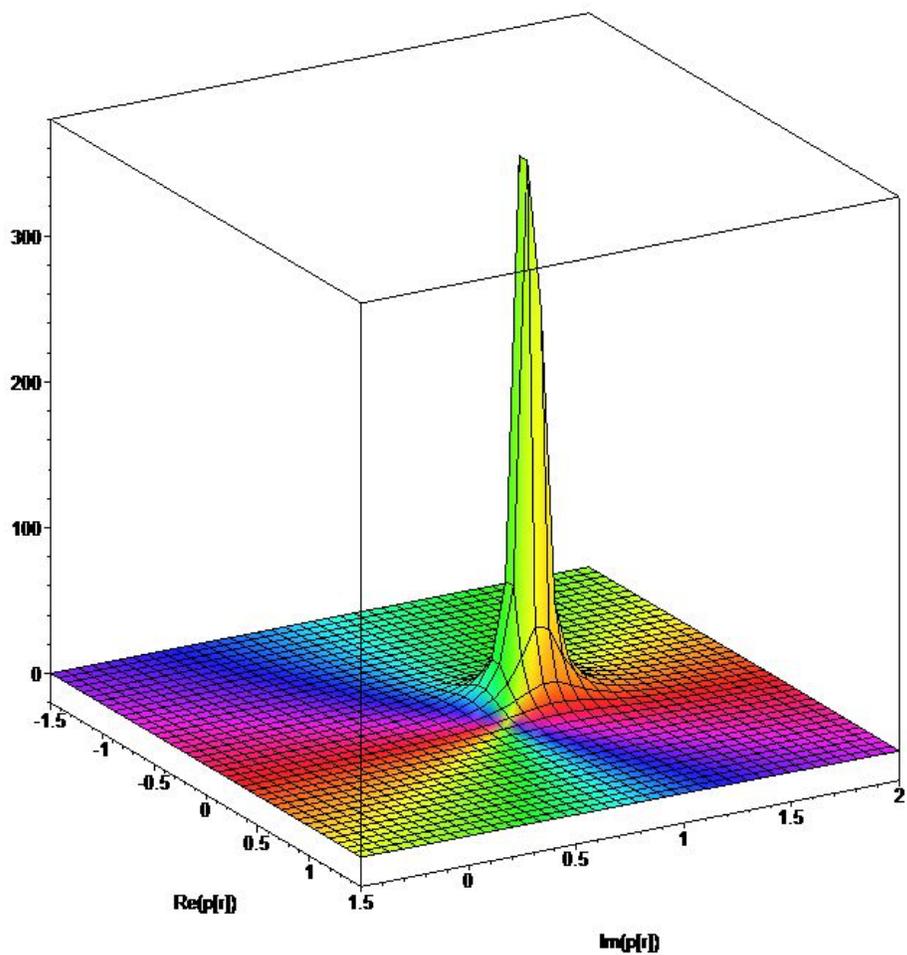

**Figure A1**. Plot of the real and imaginary part of the spherical polar momentum space function of the hydrogen atom for the n = 1, ℓ = 0, m = 0 ground state.

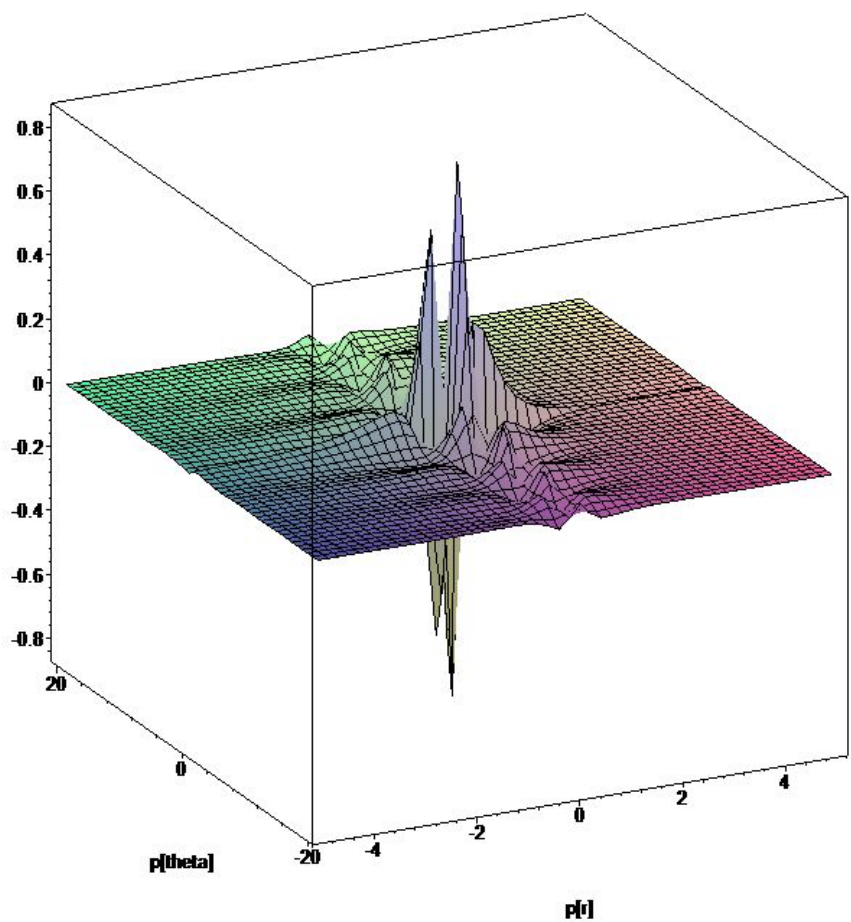

**Figure A2**. Plot of the real part of the spherical polar momentum space function of the hydrogen atom along the $(p_r, p_\theta)$ axes for the n = 2, $\ell$ = 0, m = 0 state.

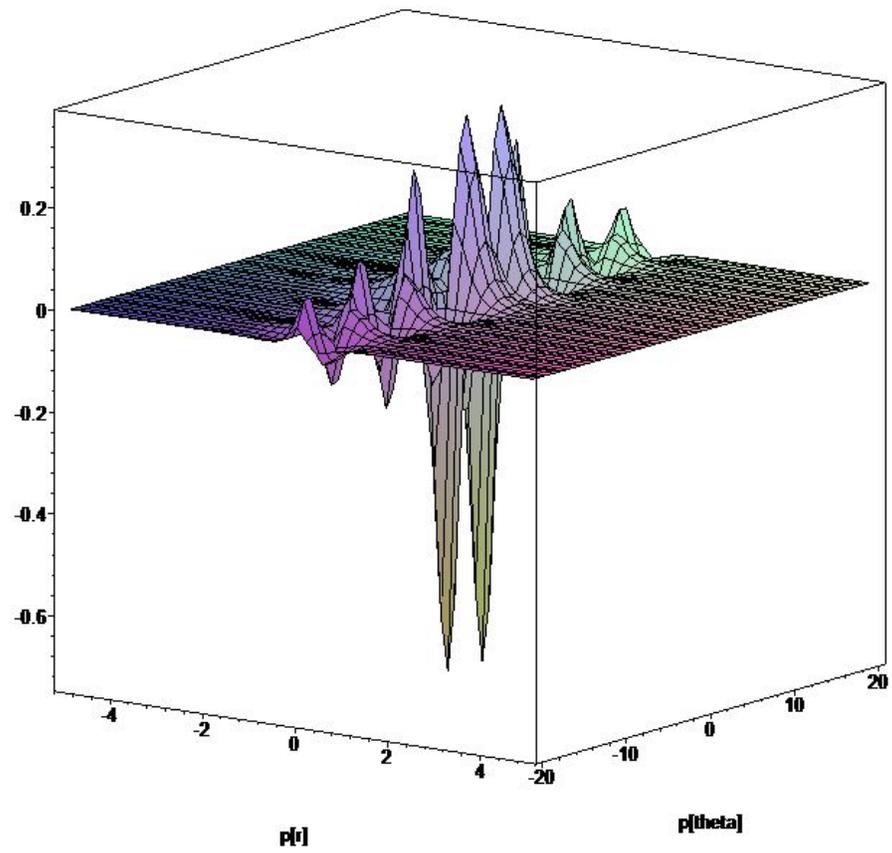

**Figure A3**. Plot of the real part of the spherical polar momentum space function of the hydrogen atom along the $(p_r, p_\theta)$ axes for the n = 2, $\ell$ = 1, m = 0 state.

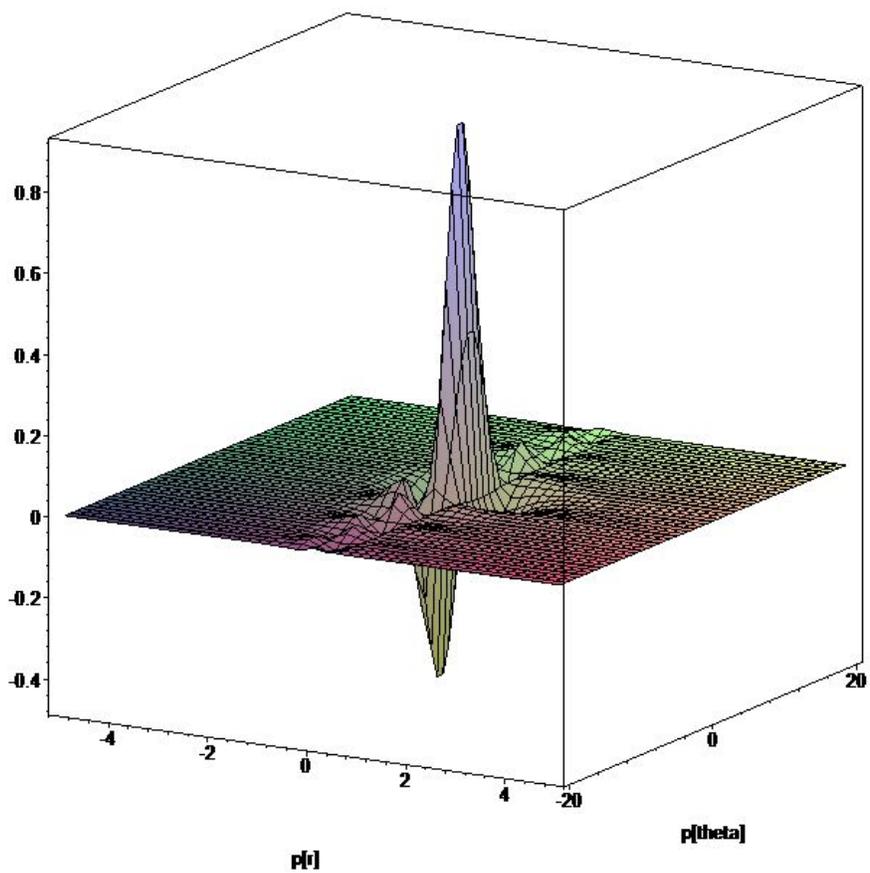

**Figure A4**. Plot of the real part of the spherical polar momentum space function of the hydrogen atom along the $(p_r, p_\theta)$ axes for the n = 2, $\ell$ = 1, m = 1 state.

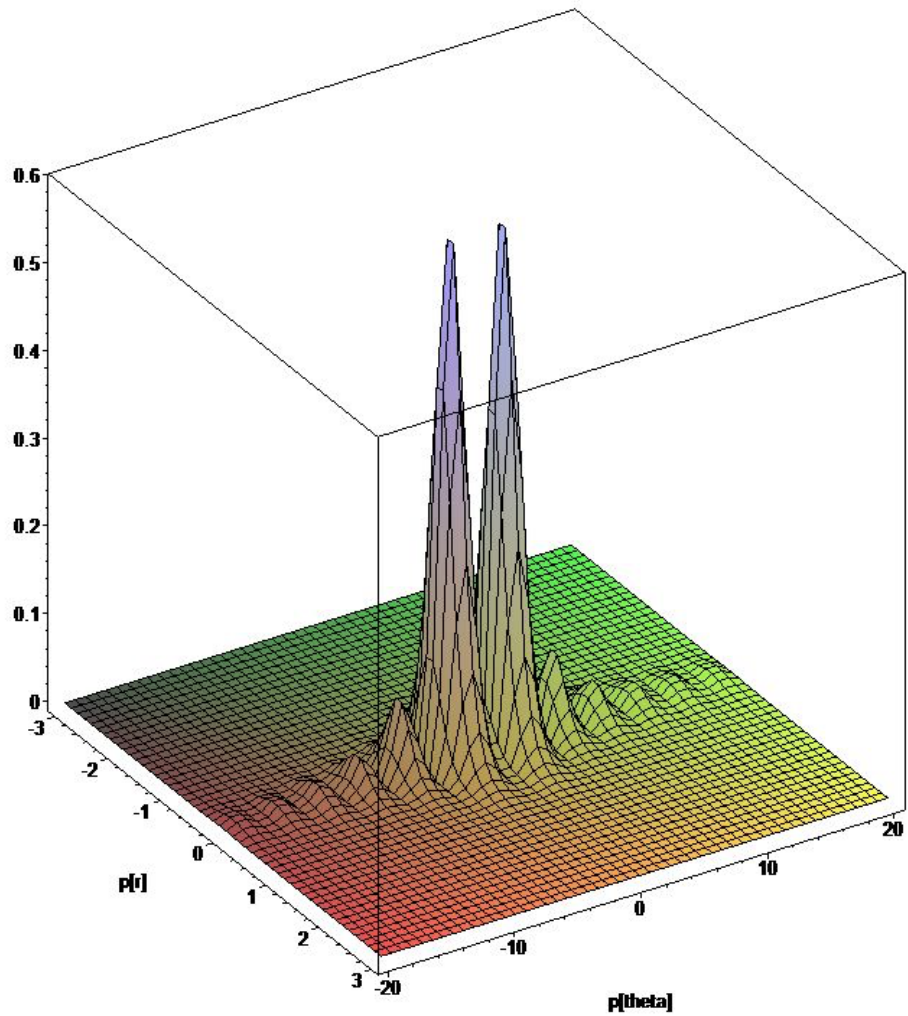

**Figure A5.** Plot of the square of the spherical polar momentum space function of the hydrogen atom along the $(p_r, p_\theta)$ axes for the n = 2, $\ell$ = 1, m = 0 state.

[16] E.T. Whittaker and G. N. Watson, *A Course in Modern Analysis*, (Cambridge University Press, 1962, Fourth Edition), Chapter VI.

[17] I.S. Gradshteyn and I.M. Ryzhik, *Tables of Integrals, Series and Products*, (Academic Press, New York, 1965, Fourth Edition), pp. 494-8, eq. 8.